\documentclass[%
 reprint,
 nofootinbib,
 nobibnotes,
 amsmath,amssymb,
 aps,mathrsfs,
 twocolumn,
 prl,
]{revtex4-1}

\usepackage{algorithm}
\usepackage{adjustbox}

\usepackage{epsfig}
\usepackage{color}
\usepackage{graphicx}
\usepackage{epstopdf}
\usepackage{footnote}
\DeclareGraphicsExtensions{.ps}
\graphicspath{{img/}}
\usepackage{subcaption}
\usepackage{float}
\usepackage{todonotes}
\usepackage{amsmath,amssymb}
\usepackage{multirow}
\usepackage{soul}

\newcommand{\bea}{\begin{eqnarray}}
\newcommand{\eea}{\end{eqnarray}}
\newcommand{\beaa}{\begin{eqnarray*}}
\newcommand{\eeaa}{\end{eqnarray*}}
\newcommand{\be}{\begin{equation}}
\newcommand{\ee}{\end{equation}}
\newcommand{\vect}[1]{\mathbf{#1}}
\newcommand{\intr}{\int\!\! d\mathbf{r}\,}
\newcommand{\intra}{\int\!\! d\mathbf{r}_1\,}
\newcommand{\intrb}{\int\!\! d\mathbf{r}_2\,}

\newcommand{\intk}{\int\!\! \frac{d\mathbf{k}}{(2\pi)^3}\,}

\newcommand{\ikrab}{\text{i}\mathbf{k}\cdot \mathbf{r}_{12}}

\newcommand{\igs}{\text{i}\mathbf{g}\cdot \mathbf{s}_{12}}
\newcommand{\ra}{\mathbf{r}_1}
\newcommand{\rb}{\mathbf{r}_2}

\newcommand{\sab}{\mathbf{s}_{12}}
\newcommand{\rab}{\mathbf{r}_{12}}
\newcommand{\ve}{V^\text{ext}}
\newcommand{\drho}{\Delta\rho}
\newcommand{\te}{\text{ex}}

\newcommand{\calF}{ {\cal F} }

\newcommand{\bfr}{\mathbf{r}}

\newcommand{\bfg}{\mathbf{g}}

\newcommand{\nvac}{n_{\mbox{\tiny vac}}}

\newcommand{\tb}{\textcolor{blue}}

\newcommand{\mf}{}
\newcommand{\mff}{}

\begin{document}

\title{The direct correlation function of \mf{a crystalline} solid}


\author{S.-C. Lin}
 \email{shang-chun.lin@uni-tuebingen.de}
\affiliation{%
 Institut 
 f\"ur Angewandte Physik, Universit{\"a}t T{\"u}bingen, Auf der Morgenstelle 10, 72076  T{\"u}bingen, Germany
}%

\author{M. Oettel}
\affiliation{%
 Institut 
 f\"ur Angewandte Physik, Universit{\"a}t T{\"u}bingen, Auf der Morgenstelle 10, 72076  T{\"u}bingen, Germany
}%


\author{J. M. Häring}
\affiliation{%
 	Fachbereich f{\"u}r Physik, Universit{\"a}t Konstanz, 78457 Konstanz, Germany
}%

\author{R. Haussmann}
\affiliation{%
 	Fachbereich f{\"u}r Physik, Universit{\"a}t Konstanz, 78457 Konstanz, Germany
}%

\author{M. Fuchs}
\affiliation{%
 	Fachbereich f{\"u}r Physik, Universit{\"a}t Konstanz, 78457 Konstanz, Germany
}%


\author{G. Kahl}
\affiliation{%
 	Institut f{\"u}r Theoretische Physik, TU Wien, 1040 Vienna, Austria
}%

\date{\today}

\begin{abstract}

Direct correlation functions (DCFs), linked to the second functional derivative
of the free energy with respect to the one-particle density, play a
fundamental role in a statistical mechanics description of
matter. This holds in particular for the ordered phases: DCFs contain
information \mff{about the local structure including defects and encode the thermodynamic properties of crystalline solids; they open a route to the elastic constants beyond low temperature expansions.}
Via a numerical {\it tour de force} we have explicitly calculated for the first time the DCF of a
solid: based on the fundamental measure concept we provide results for
the DCF of a hard sphere crystal. We demonstrate that this function
differs at coexistence significantly from its liquid counterpart -- both
in shape as well as in its order of magnitude -- because it is dominated by vacancies. We provide evidence
that the traditional use of liquid DCFs in functional Taylor
expansions of the free energy is conceptually wrong and show that the
emergent elastic constants are in good agreement with simulation-based
results.

\end{abstract}

\maketitle

\
%
%
%
%
%
%
\textit{Introduction.--}
\text
In classical and quantum theories of many--body systems, two--point correlation functions or propagators play a very important role. In homogeneous systems, they describe the fundamental structural correlations; they can be interpreted as the probability of finding two particles at two different points (in general, these are points in space and time). In functional formulations of many--body theory, these two--point functions are generically related to functional derivatives of a generating functional with respect to two local source terms. In classical systems in equilibrium \cite{evans1979nature}, the generating functional may be taken \mf{as} the grand potential $\Omega$ and the source term is the local chemical
potential defined by $\psi(\vect r)=\beta\mu-\beta\ve(\vect r)$, where $\beta=1/(kT)$ is the inverse temperature, $\mu$ is a bulk chemical potential and $\ve(\vect r)$ is an external potential acting on particles at space point $\vect r$. The corresponding second derivative
$-\beta \delta^2 \Omega/(\delta \psi(\vect r_1) \delta \psi(\vect r_2) )=G(\vect r_1,\vect r_2)$ is the total
pair correlation function. Upon a Legendre transform to a free energy $\calF[{\rho(\bfr)}]$ with the one--particle density $\rho(\vect r)$ as its natural source term variable, another correlation function may be defined by
$\beta \delta^2 {\cal F}/(\delta \rho(\vect r_1) \delta \rho(\vect r_2) )=C(\vect r_1,\vect r_2)$.
{$C(\vect r_1,\vect r_2)=C^\text{id}(\vect r_1,\vect r_2)-c(\vect r_1,\vect r_2)$ is commonly split into a trivial ideal gas part ($C^\text{id}(\ra,\rb)= \delta(\ra-\rb)/\rho(\ra)$) and an excess part,
the latter of which is called the direct correlation function DCF.}
This function is more fundamental than $G$ in the sense, that $G$ may be built by a sequence of the DCF's through the (inhomogeneous) Ornstein--Zernike relation. Knowing $c(\vect r_1,\vect r_2)$ for the stable phases or aggregate states of classical systems thus entails knowing the structural order of these phases and constitutes a desirable scientific asset. In past decades, the total pair and direct correlations of simple and complex \textit{liquids} have been studied in detail and qualitative and quantitative aspects of them are known \cite{hansen2013theory}. In contrast, this is not the case for the \textit{crystalline} state whose ordered nature is frequently only characterized by the periodicity in $\rho(\vect r)$ (which is the
first derivative $-\beta \delta\Omega/\delta \psi(\vect r)$). {Here, we aim to close this knowledge gap on the DCF and demonstrate that the shape of a  crystal DCF is 
very different from a liquid DCF and, in particular, is divergent in the limit of an ideal, defect-free crystal. } 
{Furthermore, we analyze (generalized) elastic constants, \mf{viz.~thermodynamic derivatives with respect to density and strain}, in terms of the crystal DCF and show that it encodes the mechanical properties. 

{\textit{Basic concepts.--} The appropriate functional expansion of the excess part 
(over ideal gas) of the free energy around a reference bulk state with density $\rho_0$ is given by \cite{evans1979nature}:
\begin{align}
  \label{eq:ftaylor}
  {\cal F}^\te[\rho] &= F^\te(\rho_0) + \intr \mu^\text{ex}(\vect r;\rho_0) \drho(\vect r)   \nonumber \\
   &    -\frac{1}{2\beta} \intra \intrb c (\ra,\rb;\rho_0) \drho(\ra) \drho(\rb) \nonumber \\
   & + \dots \;
\end{align}
Here, $F^\te(\rho_0)$ is the excess part of the free energy of the reference state. For a liquid, $\rho_0$ is constant and $\mu^\text{ex}$
is the constant excess chemical potential while for a crystal, $\rho_0$ and  $\mu^\text{ex}$ are lattice--periodic.} $c(\ra,\rb;\rho_0)$ is the DCF of the reference state.
In many theoretical works the classical solid is considered as a perturbation of the homogeneous liquid
{. In} Eq.~\eqref{eq:ftaylor} this amounts to approximate 
$c(\ra,\rb;\rho_0) \approx c^\text{liq}(|\ra-\rb|;\rho_0)$ with the translation invariant DCF of the liquid. 
Minimizations of such a liquid-like free energy functional can qualitatively (e.g. hard sphere systems \cite{ramakrishnan1979first}) and for some systems also quantitatively (e.g. soft systems \cite{likos2007,pini2015,mladek2006formation}) describe crystals with a periodic
$\rho(\vect r)$, the underlying assertion is, however, that the crystal DCF is liquid--like.


Earlier fundamental considerations \cite{mccarley1997correlation} cast doubts on this \mf{approach} but actual evaluations of the crystal DCF were restricted to a harmonic model.
\mf{
For a density distribution $\rho(\vect r)$ with lattice periodicity, the Fourier modes are discrete,
nonzero only for reciprocal lattice vectors (RLV) $\vect g$. 
The corresponding crystal DCF $c^\text{cr}(\ra,\rb)$ is invariant with respect to a global translation by a lattice vector 
$\vect L$, i.e. $c^\text{cr}(\ra+\vect L, \rb +\vect L) = c^\text{cr}(\ra,\rb)$. 
We can define center--of--mass and relative coordinates by
$ \sab =  \gamma\ra + \gamma' \rb$ and $\rab = \ra-\rb$ (here, $\gamma=(1-\gamma')$ is arbitrary). Thus the crystal DCF possesses an expansion \cite{mccarley1997correlation,fuchs2010}
\begin{align}
	&c^\text{cr}(\ra,\rb) =  \sum_\vect{g} \intk e^{\igs} e^{\ikrab} \, \tilde c^\text{cr}_\vect{g}(\vect k),\;\label{eq:cgk}
\end{align}
which defines the RLV modes $\tilde c^\text{cr}_\vect{g}(\vect k)$ of the DCF, where $\tilde c^\text{cr}_{\vect{g}=\vect{0}}(\vect k)$ is the Fourier transform of a translationally invariant  $c(|\ra-\rb|;\rho_0)$. Eq.~\eqref{eq:ftaylor} {with $c(\ra,\rb) \approx c^\text{liq}(|\ra-\rb|)$} fails fundamentally in considering 
{exclusively} this term.}
Later work pointed out the importance of non--liquidlike parts in the DCF to explain the occurrence of liquid--fcc [face centered cubic] vs. liquid--bcc [body centered cubic] transitions \cite{bharadwaj2013correlation}. The non--liquidlike parts in a restricted expansion have been evaluated using liquid--state methods \cite{jaiswal2014}.  

\mf{There are further arguments showing that } the approximation of crystal DCF's by their liquid counterparts  \mf{should be considered} conceptually wrong. The functional definition of $C$ entails that 
$C(\vect r_1, \vect r_2)=\delta\psi(\vect r_1)/\delta \rho(\vect r_2)$ and describes the change in chemical potential
at one point upon change of density at another point. Solids of particles with repulsive cores usually have very few vacancies
(their relative concentration $\nvac \sim 10^{-4}$). It can be shown that for a bulk solid of hard spheres 
$\psi(\vect r) =\beta\mu \propto - \ln \nvac$ \cite{oettel2010free}, i.e. it diverges for an ideal, defect--free crystal. A local change in density of the crystal should mainly be affected through a change in $\nvac$ and thus we expect the value of the DCF 
$O(c^\text{cr}) \sim 1/\nvac \approx 10^4 $, about two orders of magnitude larger than the DCF of the hard sphere liquid.   
{Moreover, according to Eq.~\eqref{eq:ftaylor}}, changes in free energy upon small density changes
in the solid through deformations should be describable with just the crystal DCF at the reference point. 
This entails a relationship of all the elastic constants to $c^\text{cr}$. Therefore,
the correct nature of the crystalline DCF should also be reflected in measurable quantities.

\textit{Direct correlation function from FMT.--}
Below we analyze the crystal DCF of hard spheres with a density functional from Fundamental Measure Theory (FMT), currently the most accurate density functional theory available. The so-called dimensional crossover route of derivation of these FMT functionals \cite{rosenfeld1996dimensional} entails that the free energy of highly localized density profiles is exact. This is crucial for the description of crystals where the density is sharply peaked at lattice sites. 
%
%
%
%
%
%

\begin{figure*}[ht]
\begin{subfigure}{0.32\textwidth}
\centering{\includegraphics[width=\textwidth]{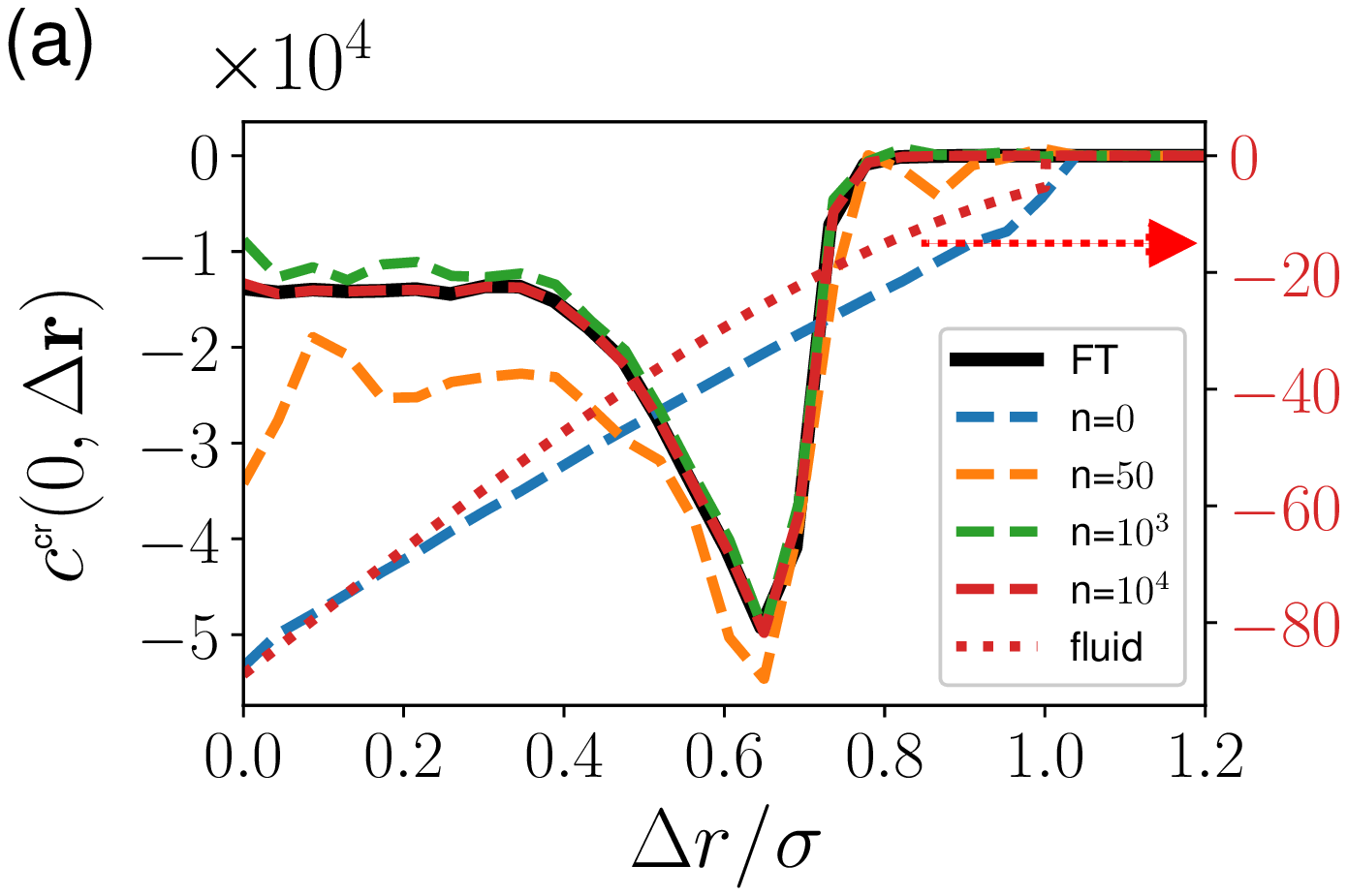}}
\end{subfigure}
\begin{subfigure}{0.32\textwidth}
\centering{\includegraphics[width=\textwidth]{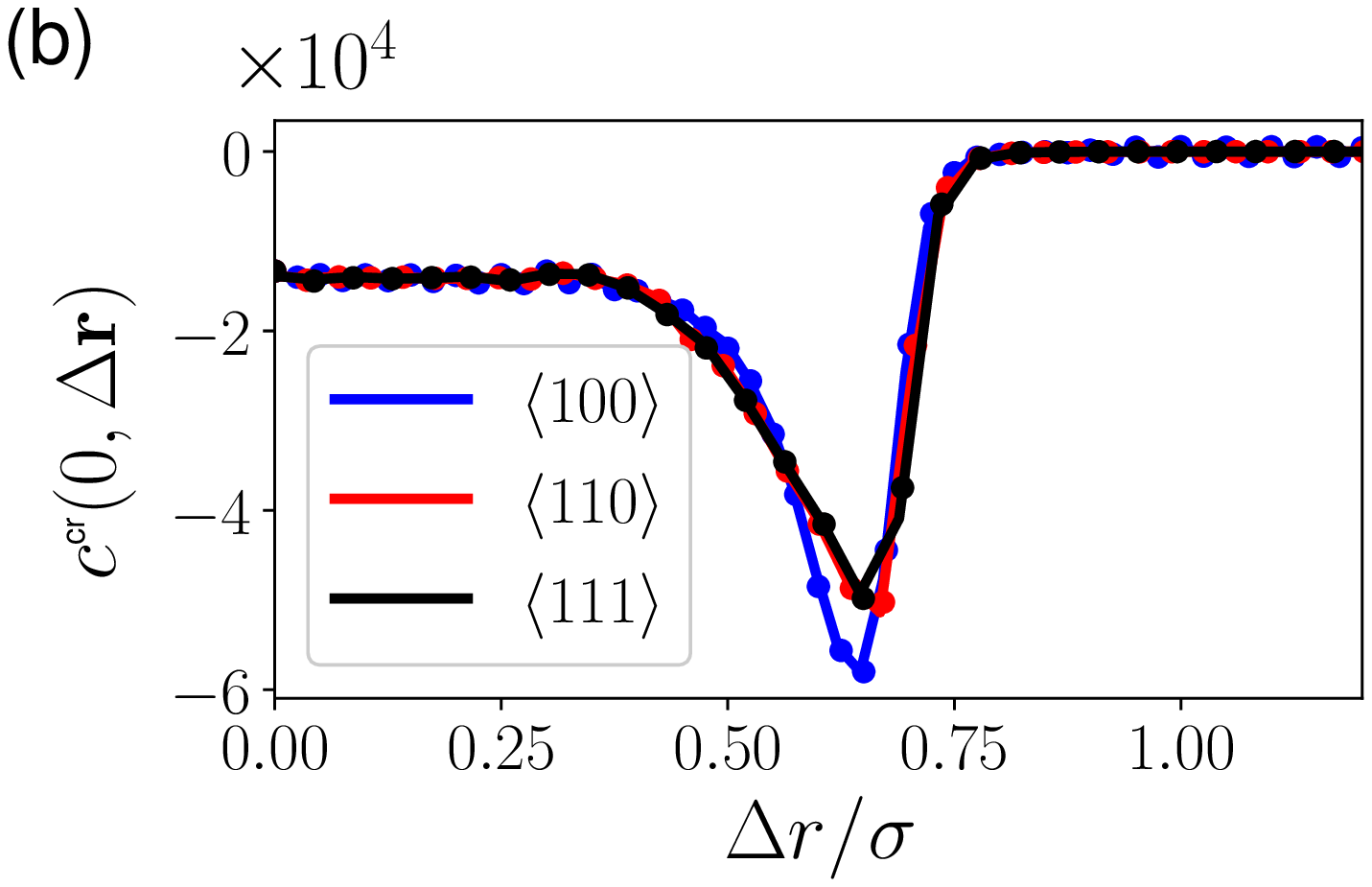}}
\end{subfigure}
\begin{subfigure}{0.32\textwidth}
\centering{\includegraphics[width=\textwidth]{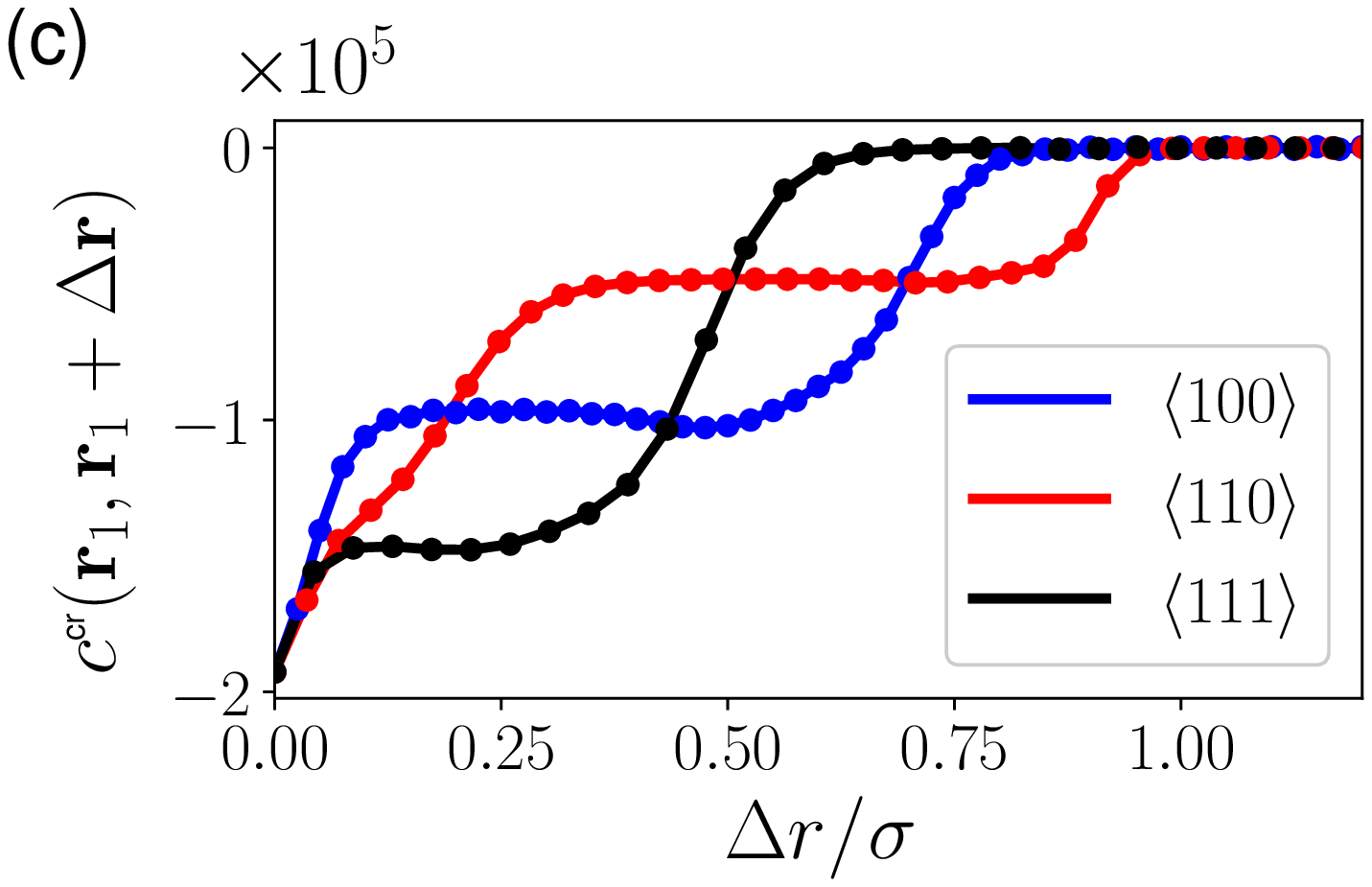}}
\end{subfigure}
\caption{Direct correlation function  $c^\text{cr}(\ra,\rb;\rho(\bfr))$ from FMT as function of distance (with $\sigma$ the sphere diameter) in a hard sphere fcc crystal at the melting density ($\eta=0.545$); the vacancy concentration is $\nvac=2.18\times10^{-5}$.  (a) $c^{\text{cr}}$ in the RLV expansion up to $10^4$ shells\mf{; $\Delta r$ is started from a lattice site}. The black solid line is from the brute force FT and the red dotted line is the fluid DCF \cite{oettel2012description} at $\eta=0.545$ with scale on the right side. (b) $c^{\text{cr}}(0,\Delta\bfr)$ in three directions by the brute force FT (solid line) and RLV expansion up to $10^4$ shells (dots). (c)  $c^{\text{cr}}(\ra,\ra+\Delta\bfr)$ with \mf{distance measured from the interstitial point} $\ra=(\frac{a}{4},\frac{a}{4},\frac{a}{4})$ and $a$ the side length of the cubic unit cell, lines are as in (b).}
\label{fig:c2_mode}
\end{figure*}

%
%
%
%
%
%


In FMT, the excess free energy is described by a local functional in a set of weighted densities (labeled by
$\alpha$),
$n_\alpha(\vect r) = \intra \rho(\vect r - \ra) w_{\alpha}(\ra)$ with $w_{\alpha} (\vect r)$ being a  corresponding weight function, defined by geometrical characteristics of the particles \cite{rosenfeld1989free}:
\bea
  \beta {\cal F}^\te[\rho] = \intr \Phi(n_\alpha(\vect r))\;,
\eea 
Here we use the White-Bear-II (WBII) tensorial version \cite{tarazona2000density,roth2010fundamental} for $\Phi$ (the full definition is given in the SI). 
Crystal density distributions, the equation of state of both liquid and crystal and consequently the 
coexistence densities of the liquid and solid are in excellent agreement with simulations \cite{oettel2010free}.
The crystal DCF follows from the second functional derivative of ${\cal F}$ with respect to $\rho$
\bea
  c^\text{cr}(\ra,\rb) &=& - \sum_{\alpha\beta} \intr \frac{\partial^2 \Phi(\vect r)}{\partial n_\alpha \partial n_\beta} \, 
      w_{\alpha}(\vect r - \ra) w_{\beta}(\vect r - \rb) \nonumber \;,\\
      \label{eq:cfmt}
\eea
%
%
%
%
%
%
\mf{which can most conveniently be analyzed in reciprocal space; see the SI.} 
\mf{We} determine $c^\text{cr}$ \mf{by two different ways, on the one hand from}  the RLV expansion of an fcc lattice and\mf{, on the other hand} (as a cross--check), via brute force \mf{six-dimensional Fourier} transformation, both   at the WBII melting point: packing fraction $\eta=0.545$ and $\nvac=2.18\times10^{-5}$. 



Results are shown in Fig.~\ref{fig:c2_mode}. 
Panel (a) shows $c^\text{cr}(0,\Delta\bfr)$, where the first point (origin) 
is a lattice point and $\Delta\bfr$ points in $\langle 111 \rangle$ direction of the fcc cubic unit cell.
\mf{The number $n$ of RLV shells considered (see the legend)} demonstrates the slow convergence of the RLV expansion (we used up to $n=10^4$ RLV shells) 
towards the result obtained from the brute force FT. The \mf{translationally invariant} RLV mode $\tilde c^\text{cr}_{\bfg=0} (n=0)$ {describes the direct correlations with the center-of-mass variable averaged over the unit cell}. It  has a similar
shape as the fluid DCF  but is more than a factor \mf{500} larger (note the separate axis scale for
the fluid DCF in Fig.~\ref{fig:c2_mode}(a)). The shape of the full result is however very different from the fluid DCF.
Fig.~\ref{fig:c2_mode}(b) shows $c^\text{cr}(0,\Delta\bfr)$ using the RLV expansion and the brute force FT in the directions of $\langle 100 \rangle$, $\langle 110 \rangle$ and $\langle 111 \rangle$.
These results demonstrate that the DCF is fairly isotropic around a lattice point.
In contrast, the isotropy is lost if an interstitial point is chosen as the first point,
see Fig.~\ref{fig:c2_mode}(c) which 
shows $c^\text{cr}(\ra,\ra+\Delta\bfr)$ in the three directions with $\ra=(\frac{a}{4},\frac{a}{4},\frac{a}{4})$ ($a$ is the side length of the cubic unit cell). Overall, the order of magnitude for $O(c^\text{cr})\sim 10^4$ agrees very well with the
estimate $1/\nvac$ from above, and the 3D spatial dependence is very different from a liquid DCF.

%
%
%
%
%
%

\textit{Elastic constants.--} 
{
The calculation of macroscopic properties of a crystal from first principles requires a correct description of the microscopic structure. Having done the first step
we can explore the ramifications of the discussed DCF properties for the elastic constants
{(we call this the DCF route to elastic constants)}. 
}
This has been done so far by using the fluid DCF only 
\cite{Lipkin1985,Jaric1988,Mahato1991,Tosi1994}.
The \mf{familiar} elastic constants are defined through an expansion of the free energy $F(\eta)$ of a strained crystal to second order in the Lagrangian strain tensor
$\eta_{\alpha\beta}=(u_{\alpha\beta}+u_{\beta\alpha}+u_{\gamma\alpha} u_{\gamma\beta})/2$ where $u_{\alpha\beta}$ is the usual gradient of the displacement field ($u_{\alpha\beta}=\nabla_\beta u_\alpha$) and like indices are summed over  \mf{\cite{Wallace1970}}
\bea
 \label{eq:defC}
\frac{F(\eta)}{V} \approx \frac{F(0)}{V} - p \; \eta_{\alpha\beta}\delta_{\alpha\beta} + \frac{1}{2}C_{\alpha\beta\gamma\delta}\;\eta_{\alpha\beta}\eta_{\gamma\delta}\;. 
\eea
Here, $V$ is the volume of the unstrained equilibrium reference state with pressure $p$ and the number of particles is fixed. 
For an fcc crystal, there are only 3 independent elastic constants which in Voigt notation are
$C_{11} = C_{\alpha\alpha\alpha\alpha}$, $C_{12} = C_{\alpha\alpha\beta\beta}$, $C_{44} = C_{\alpha\beta\alpha\beta}$ (no summation and $\alpha \neq\beta$).     Since a free energy change to second order in strain is equivalent to a change in second order in the density profile (see Eq.~\eqref{eq:ftaylor}), the elastic constants should be expressible with just the density profile and the DCF at the reference state. For the change in density $\delta\rho(\bfr)$ upon applying a linear displacement field $\mathbf{u}(\bfr)$ and a change in average
density $\delta\bar\rho(\bfr)$ one may write
\mf{\cite{Szamel1993}} 
\begin{equation}
\delta \rho(\bfr) \approx - \mathbf{u}(\bfr)\cdot\nabla\rho(\bfr) + \rho(\bfr)\frac{\delta \bar{\rho}(\bfr)}{\bar{\rho}}\;.
\label{eqn:delta_rho}
\end{equation}
\mf{While the density profile $\rho(\bfr)$ varies rapidly on the length scale of the lattice spacing, the coarse grained displacement and average density field only exhibit smooth variations.  }
This \mf{ansatz} corresponds to an affine deformation of the crystal density profile. The change in average
density $\bar\rho$ is the sum of two effects: the change in vacancy concentration (or occupancy of the unit cell) and the change of the unit cell volume {\cite{Martin1972,Fleming1976}}. 
Using this affine approximation, Refs.~\cite{fuchs2010,fuchs2015} decomposed the second order change in total free energy
 \bea
&&\Delta {\calF}^{(2)}=\frac{1}{2\beta} \intra \intrb \left(  \frac{\delta(\ra-\rb)}{\rho(\ra)}-c^\text{cr}(\ra,\rb) \right)\nonumber\\      
&&\hspace{1.4cm}\times\delta\rho(\ra) \delta\rho(\rb)\nonumber\\ \label{eq:eq7}
&&=\frac{1}{2}\int\!\!d\bfr\left(\lambda_{\alpha\beta\gamma\delta}  {u}_{\gamma\alpha} {u}_{\delta\beta} -2 \mu_{\alpha\beta} \frac{\delta\bar{\rho}}{\bar{\rho}}  {u}_{\beta\alpha}+  \nu \left(\frac{\delta\bar{\rho}}{\bar{\rho}}\right)^2\right)\,.\nonumber\\
\eea
\begin{figure*}[ht]
\begin{subfigure}{0.45\textwidth}
\centering{\includegraphics[width=\textwidth]{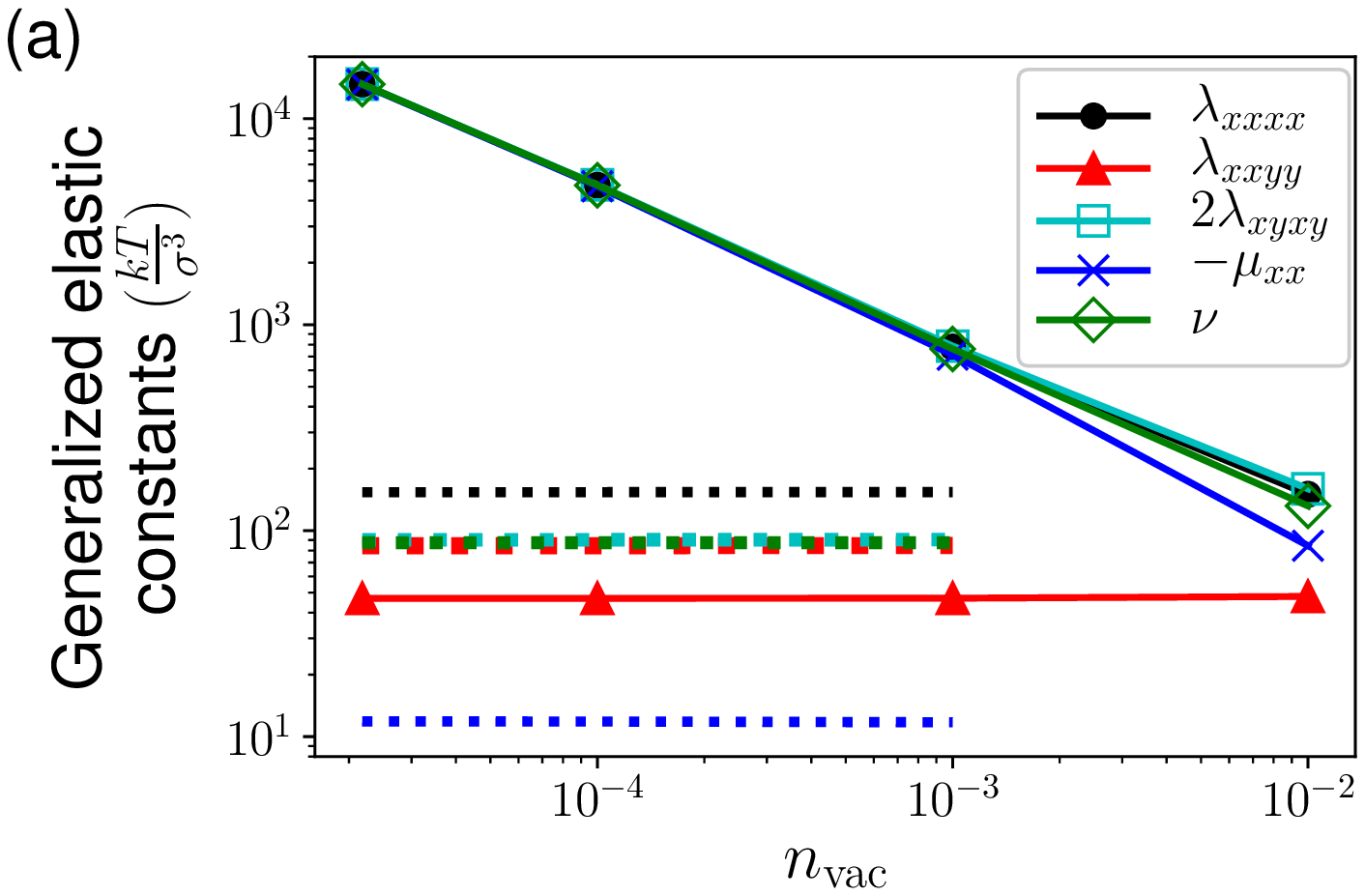}}
\end{subfigure}
\begin{subfigure}{0.45\textwidth}
\centering{\includegraphics[width=\textwidth]{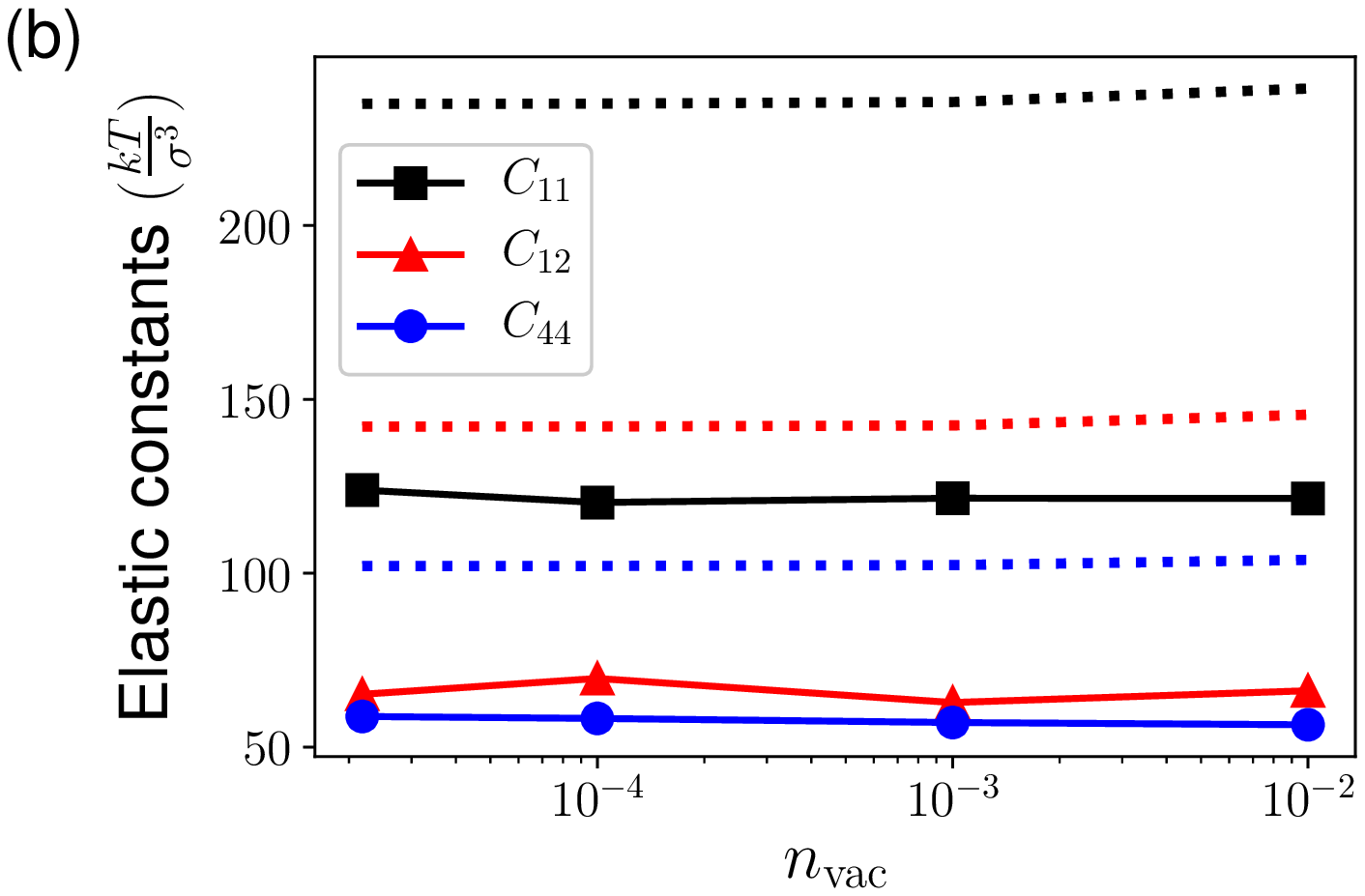}}
\end{subfigure}
\caption{ {DCF route to elastic constants: (a) Generalized elastic constants and (b) Voigt elastic constants at coexistence ($\eta=0.545$) as a function of vacancy concentration. 
Symbols connected with full lines are results from Eqs.~\eqref{eq:lmunu_def} and \eqref{eq:CVoigt}}, and 
dotted lines use Eqs.~\eqref{eq:lmunu_def} and \eqref{eq:CVoigt} with $c^\text{cr}(\ra,\rb) \to c^\text{\tiny{liq}}(|\ra-\rb|;\eta_0)$,
the fluid DCF \cite{oettel2012description} at $\eta_0=0.545$. 
}
\label{fig:elastic_const_vac}
\end{figure*}
%
This defines the generalized elastic constants
\mf{as thermodynamic derivatives}
\begin{align}
&\lambda_{\alpha\beta\gamma\delta} = \frac{1}{2} \; \hat I\left( \nabla_\alpha\rho(\ra) \, \nabla_\beta\rho(\rb)\, c^\text{cr}(\ra,\rb)\, r_{12,\gamma}\,r_{12,\delta}\right),\nonumber\\
&\mu_{\alpha\beta} = \hat I \left( \rho(\ra)\, \nabla_\alpha\rho(\rb)\, c^\text{cr}(\ra,\rb)\,r_{12,\beta}\right),\nonumber\\
&\nu = \hat I \left( \rho(\ra)\left(\frac{\delta(\bfr_{12})}{\rho(\rb)} - c^\text{cr}(\ra,\rb) \right)\rho(\rb) \right)
\label{eq:lmunu_def}
\end{align} 
where the integral operator $\hat I = \frac{1}{V\beta}\iint d\ra d\rb$ and $\bfr_{12}=\ra-\rb$ with Cartesian components $r_{12,\alpha}$. 
According to Eq.~\eqref{eqn:delta_rho}, these generalized elastic constants have the following meaning:
the $\lambda$'s are the constants sensitive to a second--order density profile variation due to a combination of ``strain--strain'', the
$\mu's$ accordingly are sensitive to ``strain--average density change'' and $\nu$ is sensitive to the combination
``average density change--average density change''. It follows that for a defect--free crystal ($\nvac \to 0$) some of the generalized coefficients must become very large since in such a case an independent variation of strain or average density would always be accompanied with a creation of interstitials whose free energy cost is large \cite{pronk2003large}.  

The \mf{standard} elastic constants in Voigt notation are a suitable combination of the generalized constants reflecting the process of
stressing the crystal 
{without the constraint of fixed density} 
{(see the SI):}
\begin{eqnarray}
    C_{11} &=& \lambda_{xxxx} +  2\mu_{xx} + \nu +p\nonumber\\
    C_{12} &=& 2\lambda_{xyxy} -\lambda_{xxyy} + 2\mu_{xx} + \nu  -p\nonumber\\
    C_{44} &=& \lambda_{xxyy} + p
    \label{eq:CVoigt}
\end{eqnarray}
with $p$ the pressure 
{Note that Eq.~\eqref{eq:CVoigt} also requires to pass from linear to Lagrange strain \cite{Wallace1970} \mff{and that the volume of the DFT integration in Eq.~\eqref{eq:eq7} differs from the volume of the reference state used in Eq.~\eqref{eq:defC}}. 

The computation of the \mf{(generalized)} elastic constants using the RLV modes \mf{of the FMT functional} 
is described in the SI. \mf{To first establish their dependence on vacancy density, we employ a constraint minimization of the FMT functional fixing $\nvac$ \cite{oettel2010free}.}
In Fig.~\ref{fig:elastic_const_vac}(a), the  generalized elastic constants are shown for different $\nvac$ at coexistence ($\eta=0.545$). Except for $\lambda_{xxyy}$, these are $\propto {1}/{\nvac}$ \mf{in agreement with our reasoning above}.
$\lambda_{xxyy}$  is comparably small and insensitive to changes in $\nvac$.  \mf{As it} describes the response to shear strains which do not \mf{generate defects}, i.e. do not create interstitials\mf{, this result also agrees with our expectation}.
 Note that
an evaluation of the generalized elastic constants by using a liquid--like DCF in Eq.~\eqref{eq:lmunu_def} gives results
almost independent on vacancy concentration; see the dotted lines in Fig.~\ref{fig:elastic_const_vac}(a). \mf{Clearly, a fluid DCF qualitatively fails to describe the thermodynamic derivatives approaching the ideal crystal limit.}

%
%
\mf{Interestingly,} the Voigt elastic constants from the DCF route (Eq.~\eqref{eq:CVoigt}, see Fig.~\ref{fig:elastic_const_vac}(b)) remain almost unchanged for $\nvac$ up to $10^{-2}$, as a result of a delicate cancellation between the generalized elastic constants which vary from $10^{4}$ to $10^{2}$. \mf{The insensitivity of the elastic constants to local defects explains why previous calculations of the}
Voigt constants \mf{using} a liquid--like DCF \mf{gave qualitatively reasonable behavior. Yet, quantitatively, the liquid DCF gives values off by} a factor of 2, see the dotted lines in Fig.~\ref{fig:elastic_const_vac}(b).

{Additionally, we compare the Voigt elastic constants from the DCF route with the ones obtained} by {an explicit} free
energy determination of suitably deformed unit cells and using Eq.~\eqref{eq:defC}.
For the choice of deformations, we follow the procedure proposed in Ref.~\cite{laird1992weighted} (see the SI for details) and perform a free minimization of the free energy without resorting to density profile parameterizations.
{In view of the high accuracy of the free energy for equilibrium crystals \cite{oettel2010free}, this should constitute a
reliable benchmark.}
{The elastic constants can be  grouped into combinations where the fcc crystal is compressed, viz.~the {b}ulk modulus $\frac 13C_{11}+\frac 23C_{12}$, and where it is not, $C_{11}-C_{12}$ and $C_{44}$.} 
In Tab.~\ref{tab:elastic_const}, {results} from {the DCF route} and from explicit deformations in FMT  are shown and compared with simulations \cite{pronk2003large} at the melting point. 
{
For all  constants, values at least two orders of magnitude smaller than the defect-dominated strain derivatives {are found}, and
{results obtained from the FMT functionals by the two different routes are rather consistent with the simulation data.} 
{Differences are largest for the {b}ulk modulus which apparently changes most during the nonaffine relaxation of the free energy; see the SI for an explicit calculation.}
The differences between the two routes point to necessary corrections to the affine approximation of the density profile change in Eq.~\eqref{eqn:delta_rho}, which is currently under investigation. }

\begin{table}[t]
\centering{
\def\arraystretch{1.2}%
\begin{tabular}{cccc} \hline \hline 
          & \hspace{5pt}DCF\hspace{5pt} & deformation(FMT)    & \hspace{5pt}simulation\hspace{5pt} \\ \hline
\multirow{3}{*}{\begin{tabular}[c]{@{}c@{}}\end{tabular}} 
 $\frac{1}{3}C_{11}+\frac{2}{3}C_{12}$ & $82.63$        & $38.4\pm 0.4$ & $37$    \\ \cline{1-4} 
  $C_{11}-C_{12}$ & $63.12$        & $61.9\pm 0.4$ & $52$  \\  \hline 
        $C_{44}$ &$58.81$        & $49.5\pm 0.5$ & $45$    \\ \hline \hline 

\end{tabular}
}
\caption{Elastic constants from the DCF route, explicit deformations (FMT) and simulation \protect\cite{pronk2003large} at the melting point. All elastic constants are defined using Eq.~\eqref{eq:defC}. For melting point, $\eta=0.543$ is used in simulations \cite{pronk2003large}, and $\eta=0.545$ with $\nvac=2.18\times 10^{-5}$ is used in FMT.} 
\label{tab:elastic_const}
\end{table}

\textit{Outlook.--} In this work, we have investigated the direct correlation function of the hard sphere solid using state--of--the--art
density functionals of FMT type. The crystal DCF is fundamentally different from the one of the hard sphere liquid \mf{as density changes require local defects, viz.~vacancies, in solids which are close to ideal.} The order of magnitude  \mf{ of the DCF is thus} proportional to the inverse vacancy concentration.
Generalized elastic constants may be defined in terms of the DCF and the crystal density profile, and show the proportionality to the inverse vacancy concentration. {Liquid--like DCF's do not entail this property.} Standard elastic constants are determined by the deformation of the lattice while defects can adjust to the strain. Thus they take finite values including in the ideal crystal limit.
{
The sensitivity of the generalized elastic constants to defect concentrations suggests
to have a closer look at defect--rich systems in the future
such as polydisperse
hard spheres (where interstitials will dominate over vacancies \cite{pronk2004}), systems of colloidal cubes \cite{smallenburg2012vacancy},  or interpolating systems to defect--dominated cluster crystals \cite{likos2007}.
}
{
We {anticipate that the generalized} constants play a role 
for solids with impurities and are relevant to  
mechanochemical coupling in such systems \cite{shi2018verifying}.
}


This work is supported by Deutsche Forschungsgemeinschaft through 
a D-A-CH grant FU 309/11-1 {and OE 285/5-1},  and the Austrian Funding Agency (FWF) under grant number I3846-N36.

\bibliography{ref}

\end{document}